\documentclass{emulateapj}

\usepackage{amsmath}
\usepackage{amssymb}
\usepackage{graphicx}
\usepackage{color}
\usepackage{ulem}

\def\gtaprx {\lower .1ex\hbox{\rlap{\raise .6ex\hbox{\hskip .3ex
	{\ifmmode{\scriptscriptstyle >}\else
		{$\scriptscriptstyle >$}\fi}}}
	\kern -.4ex{\ifmmode{\scriptscriptstyle \sim}\else
		{$\scriptscriptstyle\sim$}\fi}}}
\def\ltaprx {\lower .1ex\hbox{\rlap{\raise .6ex\hbox{\hskip .3ex
	{\ifmmode{\scriptscriptstyle <}\else
		{$\scriptscriptstyle <$}\fi}}}
	\kern -.4ex{\ifmmode{\scriptscriptstyle \sim}\else
		{$\scriptscriptstyle\sim$}\fi}}}

\newcommand{\cutt}[1]{\textcolor{blue}{}}

\newcommand{\code}[1]{{\tt{#1}}}
\newcommand{\Msun}{{\ensuremath{\mathrm{M}_{\odot} }}}

\newcommand{\C}{{\ensuremath{^{12}\mathrm{C}}}}
\newcommand{\Ni}{{\ensuremath{^{56}\mathrm{Ni}}}}

\begin{document}

\title{The Early Evolution of Primordial Pair-Instability Supernovae}

\author{C.C. Joggerst\altaffilmark{1,2} and Daniel
J. Whalen\altaffilmark{3}}

\altaffiltext{1}{Department of Astronomy and Astrophysics,
University of California at Santa Cruz, Santa Cruz, CA 
95060. Email:cchurch@ucolick.org}

\altaffiltext{2}{Nuclear and Particle Physics, Theoretical Astrophysics 
and Cosmology (T-2), Los Alamos National Laboratory, Los Alamos, NM 87545}

\altaffiltext{3}{McWilliams Fellow, Department of Physics, Carnegie Mellon 
University, Pittsburgh, PA 15213}

\begin{abstract}

The observational signatures of the first cosmic explosions and their chemical
imprint on second-generation stars both crucially depend on how heavy elements 
mix within the star at the earliest stages of the blast.  We present numerical 
simulations of the early evolution of Population III pair-instability supernovae 
with the new adaptive mesh refinement code \code{CASTRO}.  In stark contrast to
15 - 40 M$_{\odot}$ core-collapse primordial supernovae, we find no mixing in 
most 150 - 250 M$_{\odot}$ pair-instability supernovae out to times well after 
breakout from the surface of the star.  This may be the key to determining the 
mass of the progenitor of a primeval supernova, because vigorous mixing will 
cause emission lines from heavy metals such as Fe and Ni to appear much sooner 
in the light curves of core-collapse supernovae than in those of pair-instability 
explosions. Our results also imply that unlike low-mass Pop III supernovae, whose 
collective metal yields can be directly compared to the chemical abundances of 
extremely metal-poor stars, further detailed numerical simulations will be 
required to determine the nucleosynthetic imprint of very massive Pop III stars 
on their direct descendants.

\end{abstract}

\maketitle

\section{Introduction}

The first stars in the universe form at $z\sim$ 20 and are likely very massive, 
30 - 500 M$_{\odot}$ \citep{bcl99,abn00,abn02, bcl02,nu01,on07}.  The fates of
these stars depend on their masses: 15 - 50 M$_{\odot}$ stars die in core 
collapse supernovae (CC SNe), 140 - 260 M$_{\odot}$ stars explode in far more 
energetic thermonuclear pair-instability supernovae \citep[PISNe][]{hw02}, and 
40 - 60 M$_{\odot}$ stars may die as hypernovae, whose explosion mechanisms are 
not yet understood but are thought to have energies intermediate to those of CC 
and pair-instability SNe \citep{Iwamoto2005}. Most Pop III SNe \citep{byh03,ky05,
get07} occur in low densities (0.1 - 1 cm$^{-1}$) because UV radiation from the 
star sweeps most of the baryons from the dark matter halo in which it resides 
\citep{wan04,ket04,abs06,awb07,wa08a}. 

Metals from Pop III SNe determine the character of second-generation stars 
and the primeval galaxies they populate by enhancing cooling in the gas in 
which such stars form, and therefore the mass scales on which it fragments.  
The manner in which the first metals contaminate pristine gas in the 
primordial IGM crucially depends on mixing processes that have only begun 
to be studied numerically.  Preliminary calculations indicate that metals 
from Pop III SNe mix with gas in a halo on two characteristic spatial 
scales prior to their emergence into cosmological flows on kpc scales 
\citep{wet08a}: 10 - 15 pc, when a reverse shock forms in the remnant, and 
150 - 200 pc, when the remnant collides with the dense shell of the relic H 
II region of the progenitor.  New simulations now prove that violent mixing 
can occur \textit{within} the star prior to shock breakout from its surface, 
if the star is 15 - 40 M$_{\odot}$ \citep[][hereafter JET10] {jet09b}.
Capturing mixing on the smallest scales is prerequisite to following its 
cascade out to the largest ones, and hence to determining the colors and 
morphologies of primitive galaxies. Early mixing also governs which elements 
are imprinted on next-generation stars, whose chemical abundances may impose 
indirect constraints on the masses of the stars that enriched them. Low-mass 
remnants of this generation are now sought in ongoing surveys of ancient, 
dim metal-poor stars in the Galactic halo \citep{bc05,fet05}.  Since mixing 
in Pop III SNe in part sets elemental abundances in second-generation stars, 
it is integral to future measures of the primordial IMF.
  
Early mixing is also key to computing the light curves and spectra of the first 
cosmic explosions, whose detection could yield the first \textit{direct} measure 
of the primordial IMF. Optical and UV radiation breaks free of the SN shock when 
the ejecta reaches the outer envelope of the star and is exposed to the IGM.  In 
the frame of the shock, the photosphere from which photons escape into space 
descends deeper into the ejecta as it expands outward because of the spherical 
dilution of the ejecta. If mixing precedes radiation breakout, it determines the 
elements that the photosphere encounters as it sinks deeper into the ejecta, and 
therefore the emission lines that propagate into the IGM over time.  Accurate 
multidimensional models of the explosion from its earliest stages are therefore 
necessary to compute lines in primordial SN light curves and spectra.

Mixing in galactic core-collapse supernovae has been studied for over twenty 
years, particularly in connection with SN 1987A \citep[e.g.][and references 
therein]{fryer07}.  A core objective of these studies is to understand the 
premature appearance of \Ni\ in the spectra of 1987A, whose emission lines 
appear much earlier than a simple picture of segregated, spherically-symmetric 
expanding mass shells would predict.  Mixing in Pop III core-collapse SNe was 
first examined by \citet{jet09a}, who found that both vigorous mixing and 
fallback onto the compact remnant in 15 - 40 M$_{\odot}$ Pop III SNe govern 
which metals escape into the IGM at high redshifts.

PISNe may have been commonplace in the primeval universe, but their enrichment 
of the early IGM is yet to be understood.  To investigate the propagation of 
metals into the IGM by such explosions, constrain their nucleosynthetic imprint 
on second-generation stars, and to evaluate the impact of mixing on PISN spectra, 
we have performed two-dimensional simulations of the explosions of 150 - 250 M$_{
\odot}$ stars.  In $\S \, 2$ we review how blast profiles from the \code{KEPLER} 
code were ported as initial conditions to \code{CASTRO} and discuss what factors
govern the presupernova structure of the star.  We examine the outcomes of the 
explosions in $\S \, 3$ and in $\S \, 4$ we conclude.

\section{Models}

\subsection{\code{KEPLER}}

As in JET10, the simulations in our PISN survey were carried out in two stages.  
First, primordial stars were evolved through all stages of stable nuclear burning 
from the zero-age main sequence to initial collapse via the pair instability in the 
one-dimensional Lagrangian stellar evolution code \code{KEPLER} \citep{Weaver1978,
Woosley2002}.  The PISN is triggered when this collapse incites explosive O, and 
some cases Si, burning.  Unlike the simulations of JET10, in which the blast is 
artificially launched with a piston, the pair instability and subsequent collapse
that triggers these explosions are an emergent feature of the stellar evolution 
calculation.  They are genuinely spherical, barring (magneto)rotational effects, 
and their energies are set by O and Si burning.  The blast was followed until the 
end of all nuclear burning, $\sim$ 20 s after the start of the explosion.  The 
energy generated was computed with a 19-isotope network up to the point of oxygen 
depletion in the core of the star and with a 128-isotope quasi-equilibrium network 
thereafter.  

\begin{deluxetable*}{lcccccc}  
\tabletypesize{\scriptsize}  
\tablecaption{PISNe Models: Properties at time models were mapped to 2D \label{tab:table1}}
\tablehead{
\colhead{model} & \colhead{$M_{He}$ (\Msun)}& \colhead{$M_{N}$ (\Msun)}& \colhead{$M_{\Ni}$ 
(\Msun)}& \colhead{$M_{final}$ (\Msun)}  & \colhead{$R$ ($10^{13}$ cm)}& \colhead{$E_{kin}$ 
($10^{51}$ ergs)}}
\startdata 
u150 &  41 &  9.1(-5) &  0.079  & 143   & 16    &   3.7 \\
u175 &  46 &  1.1(-4) &  0.72   & 164   & 18    &   9.5 \\
u200 &  50 &  1.3(-4) &  5.2    & 183   & 19    &  17   \\
u225 &  54 &  1.1(-3) & 17      & 200   & 33    &  28   \\
u250 &  68 &  1.7(-4) & 38      & 238   & 23    &  34   \\
z175 &  53 &  1.6(-5) &  0.24   & 175   &  4.2  &   4.7 \\
z200 &  60 &  1.8(-5) &  2.0    & 200   &  4.5  &  12   \\
z225 &  67 &  1.7(-5) &  8.8    & 225   &  4.9  &  17   \\
z250 &  74 &  1.6(-5) & 23      & 250   &  6.2  &  29   \\
\enddata
\label{tab:table}
\end{deluxetable*}  

\subsection{\code{CASTRO}}

The one-dimensional explosion profiles were then mapped onto two-dimensional $RZ$ 
axisymmetric grids in \code{CASTRO} (Compressible ASTROphysics), a multi-dimensional 
Eulerian AMR code with a high-order unsplit Godunov hydro solver \citep{Almgren2010}. 
Each explosion was evolved past breakout from the surface of the star until all 
mixing ceased and each element in the ejecta was expanding homologously in mass 
coordinate.  We smoothly join the density at the surface of the star to a uniform 
circumstellar medium of 1 cm$^{-3}$ with an $r^{-3.1}$ power law, in keeping with 
the usual assumption of a low-density H II region around the progenitor star with no 
wind-blown shell or prior mass ejection.  The medium beyond the star has no effect 
on the dynamics within the star if its density falls more steeply than $r^{-3}$.  In 
mapping the radial profile onto the $RZ$ grid in \code{CASTRO}, care was taken to 
resolve the key elements of the explosion: the shock, the shells of elements, and the 
high-density core.  In particular, both the \Ni \ core and the O shell were resolved 
with a minimum of 16 cells.  Our mapping excludes departures from spherical symmetry 
due to O burning, but such perturbations would likely be high mode and low amplitude 
and therefore have minimal effect on the evolution of instabilities.  Our models thus 
only capture later asymmetries of mode greater than $l=1$ or 2. 

As in JET10, we adopt a monopole approximation for self-gravity. We first compute a 
radial average of the density from the $RZ$ grid to create a one-dimensional density 
profile. We then compute a one-dimensional gravitational potential from this profile 
and map it back onto the $RZ$ grid.  Since departures from spherical symmetry in the 
densities are minor, this approximation has a negligible effect on the final state of 
the explosion.  Because the PISN completely disperses the star, there is no compact 
remnant, fallback, or thus any need to include a point potential centered at the 
origin of the coordinate mesh.

We follow 15 chemical elements as individual species, each with their own continuity 
equation, and calculate local energy deposition due to radioactive decay of \Ni \ in 
the same manner as JET10. However, the formation of a nearly degenerate core at white 
dwarf densities in the progenitor necessitates the use of the Helmholtz equation of 
state (EOS) at early stages of the explosion.  As the ejecta expands and cools, we 
transition back to the ideal EOS used in JET10, which assumes that the gas is fully 
ionized and includes contributions from both radiation and ideal gas pressure.

The base grid is 1024$^{2}$, with the initial outer boundaries set so that the inner 
portion of the star is resolved as described above using no more than 6 levels of 
refinement.  The star is centered at the lower left corner of the mesh.  We apply
reflecting and outflow boundary conditions to the inner and outer boundaries of the 
grid, respectively.  Our refinement criteria are the same as those in JET10.  When 
the shock nears the edge of the grid, the simulation is stopped, the grid is doubled, 
and the calculation is then restarted, subtracting or adding levels of refinement as 
needed.  This procedure is repeated up to 12 times, depending on the model.  We halt 
the simulation when all chemical species are expanding homologously, which always 
occurs by the time the ejecta has propagated a short distance into the uniform 
circumstellar density.

\subsection{Progenitor Models}

JET10 found that mixing in low-mass Pop III core-collapse explosions primarily 
depends on the internal structure of the progenitor.  We likewise expect the
early evolution of pair-production explosions to be determined by the envelope
of the star, which is determined its mass, internal convective mixing over its 
lifetime, and by its metallicity.  Capturing the full range of structures for 
these stars is essential to a comprehensive survey of early mixing in Pop III 
supernovae.

\subsubsection{Convective Mixing}

The initial absence of metals and the large contribution of radiation to the 
pressure in massive Pop III stars promotes convection within them.  The CNO 
cycle cannot begin in primordial stars until a threshold mass fraction of 
\C\ is first created by the triple $\alpha$ process. This trace \C\ sets the 
entropy of the hydrogen layer to be just above that in the core, without the 
sharp entropy gradient in the upper layer of the helium shell that is usually 
present in He burning stars with metals.  This plus radiation pressure 
facilitates convection. In 140 - 260 M$_{\odot}$ stars, the central convection 
zone can approach, come in contact with, or even reach into the lower hydrogen 
layers, mixing them with carbon brought up from the core during helium burning.  
When these two high-temperature components mix, they burn vigorously, elevating 
energy release rates in the H shell by up to several orders of magnitude (the
so-called hydrogen boost).  

Convection affects the structure of the star in two ways.  First, since it 
raises energy production (and hence opacities) in the lower hydrogen layers, 
the star can puff up by more than an order of magnitude in radius and acquire 
a red supergiant structure. Second, if convection is extreme the transport of 
material out of the core could reduce its size and explosion energy in 
comparison to modest convection.  Unfortunately, one-dimensional stellar 
evolution models cannot predict these inherently three-dimensional processes 
from first principles.  Instead, they parametrize them with semi-convection 
coefficients.

\subsubsection{Metallicity}

Gas in high-redshift halos that is enriched to metallicities below 10$^{-3.5}$
Z$_{\odot}$ fragments on mass scales that are essentially identical to those of 
pristine gas and still forms very massive stars \citep[e.~g.][]{bcl01,mbh03,ss07}.  
However, such small metal fractions are more than enough to enhance CNO reaction
and energy generation rates in the hydrogen burning layers of the star, enlarging
it in the same manner as convective mixing \citep{Hirschi2007,Ekstrom2008}. Hence, 
we would expect a strong degeneracy between the influence of metals and convection 
on the envelope of the star, and that the full range of mixing in PISNe can be as 
easily spanned by metallicity as by convective overshoot.  

We considered 150, 175, 200, 225 and 250 M$_{\odot}$ non-rotating progenitors at 
$Z = 0$ (the z-series) and $Z = 10^{-4} Z_{\odot}$ (the u-series), which we 
summarize in Table \ref{tab:table1}.  The $Z = 0$ 150 M$_{\odot}$ star collapses 
to a black hole without exploding, so we exclude it from our \code{CASTRO} models.  
We employ metallicity rather than convective overshoot to cover the range of 
plausible progenitor structures in our study, because there is uncertainty about 
how much semi-convection there is in a given star but none about the range of 
metallicities over which it can form ($Z$ = 0 and 10$^{-4}$.  Comparison of Table 
\ref{tab:table1} with Table 1 of \citet{sc05} confirms that for a given progenitor 
mass these two metallicities do yield upper and lower limits to stellar radius
similar to those for all reasonable values of semi-convection coefficients.

\section{Results}

\begin{figure*}
\plotone{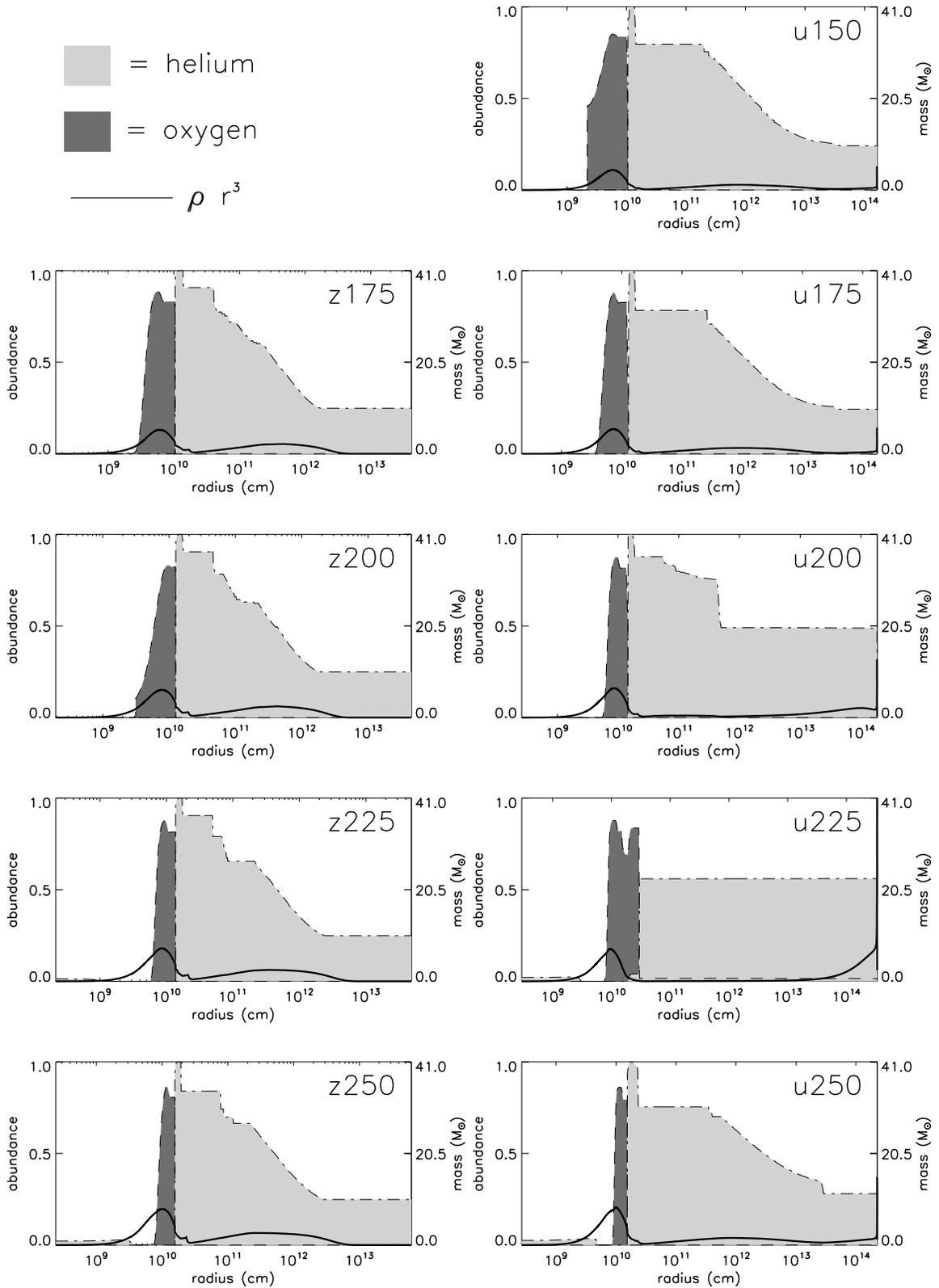}
\caption{The structure of helium and oxygen shells (delineated by
  dashed lines for clarity) in all the progenitors, with $\rho r^3$
  superimposed on them. The $\rho r^3$ profiles have all been scaled
  to the maximum $\rho r^3$ in model u225.  $\rho r^3$ increases
  dramatically near the edge of some models that die as red giants,
  while none of the blue giants show a similar structure.  The
  plateaus in helium abundance above cosmological values, which are
  most apparent in model u225, denote convective regions in which the
  hydrogen envelope mixed with a portion of the helium layer.  This
  process completely disrupted the helium layer in model u225, and
  mixed a smaller fraction of the helium layer with the hydrogen
  envelope in model u200.  A small amount of the helium shell
  convectively mixed with the hydrogen envelope in model u250.}
\label{fig:rhor3}
\end{figure*}

As expected, the u-series models die as red giants, with radii more
than an order of magnitude larger than those of the z-series, which
die as blue giants.  As shown in Table \ref{tab:table1}, the u-series
models in general have more energetic explosions than the z-series
models for a given mass.  As we show in Figure \ref{fig:rhor3},
convective mixing has completely disrupted the helium layer in model
u225 and mixed it with the hydrogen envelope.  The helium layer has
also been mixed with the hydrogen envelope, although to a lesser
extent, in model u200 and to an even slighter degree in model u250.
These u-series models also experienced some mass loss due to
pulsations.

The $\rho r^3$ values through which the shock passes are an effective predictor of
post-explosion hydrodynamics.  In regions where $\rho r^3$ increases, the shock 
must decelerate, and a reverse shock forms that inverts the pressure gradient and 
induces Rayleigh-Taylor (RT) instabilities.  Models with steeper values of $\rho 
r^3$ will experience more mixing because the forward shock slows down more abruptly, 
which leads to a stronger reverse shock. We show $\rho r^3$ values for all models
in Figure \ref{fig:rhor3}, scaled to the maximum value in model u225. It is clear 
from this figure that models u225 and u200 exhibit the largest increases in $\rho 
r^3$ near the outer edge of the star. This is due to slight bumps in density that 
are connected to the pulsations that ejected mass from these stars earlier.

\begin{figure*}
\plotone{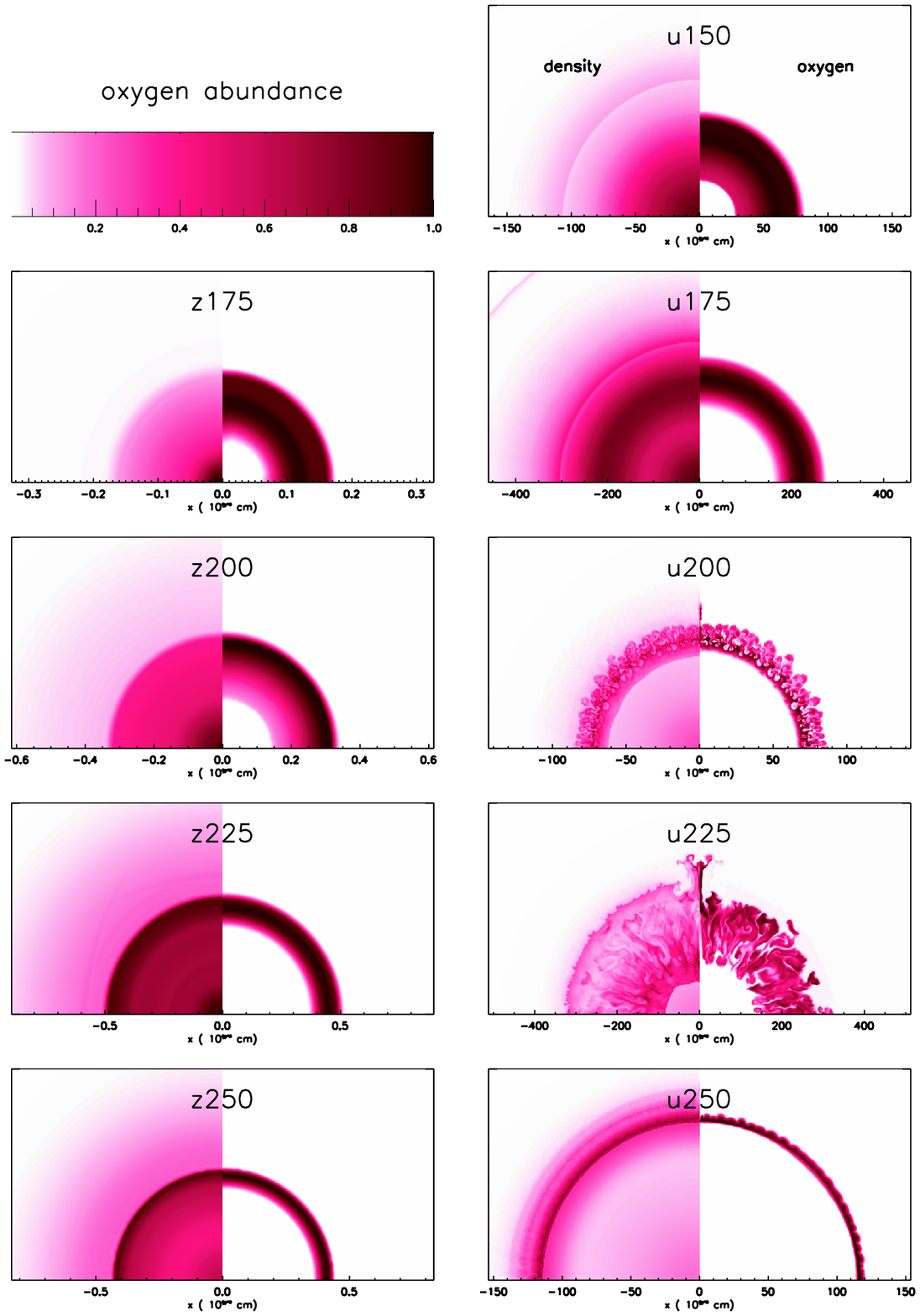}
\caption{Images of density and oxygen abundance for all models after
  the forward shock exits the star or the reverse shock (if one forms)
  traverses the star, whichever is later. Density is scaled to the
  minimum and maximum values within a given simulation.  A dense shell
  formed in all models with masses greater than 200 \Msun, but the RT
  instability only grows appreciably in the u200 and u225 models.
  These were the largest progenitors in radius and had the steepest
  $\rho r^3$ curves.  Only slight RT growth is evident in model u250.}
\label{fig:snap}
\end{figure*}

We show final states of mixing for the PISNe in Figure \ref{fig:snap}. The models 
are shown at different times, but always after the shock emerged from the surface 
of the star and the reverse shock, if one formed, dissipated. Density is shown on
the left in each panel, with values scaled to the maximum and minimum densities 
in the simulation, while oxygen is shown on the right, with the values indicated
in the color bar.  Some similarities between the models exist across all stellar 
structures. In the higher-mass compact models (z225 and z250) and in the red 
u-series models above 150 solar masses, the bulk of the oxygen layer was swept up 
into a shell of higher density than the material on either side. 

In general, however, the z-series explosions evolved quite differently than the 
u-series SNe.  In the z-series, no reverse shock formed.  A reverse shock formed
in all the u-series models, but dissipated by the time it reached the density 
contrast at the oxygen layer (u150 and u175) or was too weak by the time it 
reached this dense shell to induce rapid RT growth (u250).  In two models, a 
reverse shock formed that was strong enough to drive rapid growth of RT 
instabilities (u200 and u225).  The reverse shock was strongest in model u225, 
which had the steepest increase in $\rho r^3$ and the second highest explosion 
energy, after model u250.  The RT instability is clearly visible in the u200 
and u225 panels in Figure \ref{fig:snap}.  In these models the instability 
became nonlinear because individual RT fingers strongly interacted.  In model 
u250, the RT instability only reached the early linear phase before its growth 
was halted.

\begin{figure*}
\plotone{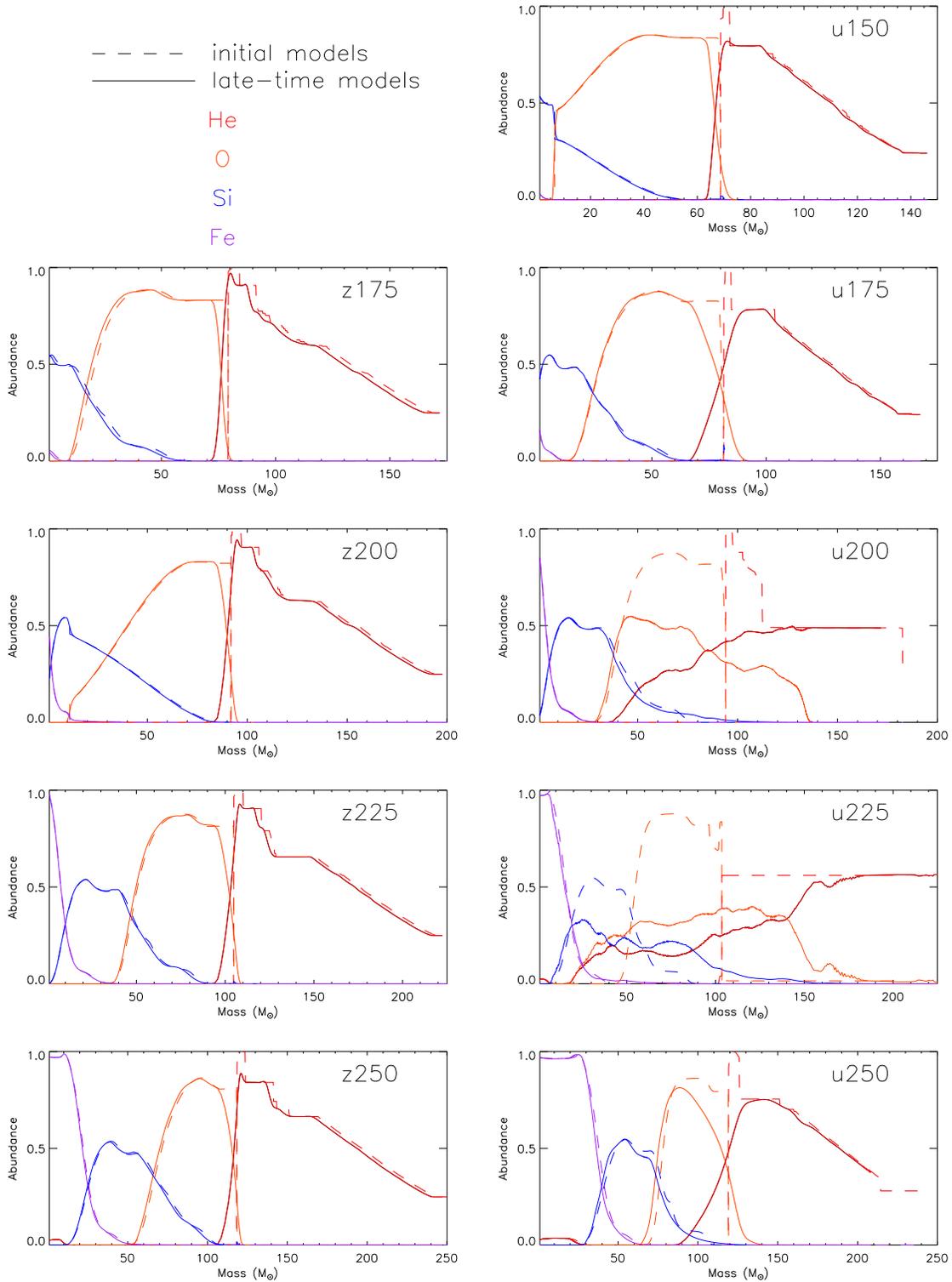}
\caption{Initial and final abundances of He, O, Si, and Fe for all models.  The 
  RT instability is only present in models u200, u225, and to a small degree in 
  u250.  The slight disagreement between initial and final profiles for models 
  which experienced no mixing is due to numerical diffusion over the evolution 
  times of these models.  In none of the mixed models did mixing break through 
  the Si layer of the star, so \Ni \ never reaches the surface of the explosion
  as it does in core-collapse supernova explosions.}
\label{fig:prof}
\end{figure*}

The extent to which the structure of the PISN at the time of explosion was
preserved or disrupted by mixing is shown in Figure \ref{fig:prof}.  We plot 
the initial shell structure of the PISN at the time of mapping into \code{CASTRO} 
as a dashed line, and overlay a solid line indicating the abundances of these 
elements after the shock exited the star or the reverse shock traversed the star, 
whichever was later.  The slight smoothing in the final profiles in comparison to 
the initial ones is due to numerical diffusion over the course of the simulation. 
The u200 and u225 panels demonstrate the extent of mixing in these explosions. The 
oxygen layer has been completely disrupted, and mixing has penetrated just to the 
top of the silicon layer (in u200) or through the silicon layer (in u225).  The 
most vigorous mixing occured in model u225, where hydrogen shell boost led to 
convection that completely mixed the helium shell and the hydrogen envelope.  The 
only other models to experience RT mixing, models u200 and u250, are also the only 
other models that show evidence of convective mixing in the outer region of the 
star prior to explosion. Model u200 manifests more RT mixing and a larger region 
(in radius and mass) in which convection mixed part of the helium layer with the 
hydrogen envelope prior to explosion.  Model u250, which has the smallest amount 
of convection extending into the helium layer, also exhibits the least RT growth. 
Models with no convective mixing between the helium layer and the hydrogen 
envelope manifest no RT mixing during the explosion, so mixing seems tied to the 
depth to which this convective envelope has penetrated the helium layer of the 
star.  The deeper the convective envelope, the more RT mixing occurs during the 
supernova shock.

RT-induced mixing in PISNe, if it occurs at all, is unlikely to leave the remnant 
in a state that resembles a core-collapse supernova remnant. In particular, it is 
unlikely to dredge up significant amounts of oxygen, let alone silicon or \Ni, 
into the outer regions of the star, or draw much hydrogen back towards the center 
of the blast.  The most vigorous mixing occurs in the model with the helium shell 
that has been completely disrupted by convective mixing with the hydrogen envelope, 
and likely represents the most mixing that can occur in a PISN. Even in this model, 
\Ni \ cannot reach the upper layers of lighter elements.

As noted earlier, our \code{CASTRO} simulations exclude the first stages of the 
explosion, in which oxygen burning drives convective mixing that may perturb the 
star \citep{chen10}.  Since these perturbations are expected to be of low mode
and high amplitude, it is unlikely they would alter the essential conclusions of 
our survey.  Reprising our two-dimensional calculations in three dimensions 
is unlikely to yield significantly different results. The RT instability initially 
grows about $30\%$ faster in three dimensions than in two because of artificial 
drag forces arising from two-dimensional geometries \citep{Hammer2010}.  However, 
once the fingers of the instability begin to interact with one another, they mix 
more efficiently in three dimensions than in two, reducing the Atwood number and 
hence their growth rate \citep{jet10}.  These two effects cancel each other, and 
the width of the mixed region in two and three dimensions is the same.  For 
spherical simulations like ours in which the RT fingers significantly interact, 
two-dimensional and three-dimensional simulations would exhibit comparable 
degrees of mixing. Also, the RT instability will not appear in three dimensions 
when it is not manifest in two.

\section{Discussion and Conclusions}
Unlike 15 - 40 M$_{\odot}$ core-collapse Pop III SNe,
150 - 250 M$_{\odot}$ Pop III PISNe experience either no internal
mixing prior to shock breakout from surface of the star or only modest
mixing between the O and He shells.  Minor mixing occurs in only two
of the explosions and is due to the formation of a reverse shock that
is strong enough to trigger the RT instability at the dense shell
created by the forward shock at the top of the oxygen layer.  The
degree of mixing is principally a function of how well hydrogen shell
boost mixes the helium shell with the hydrogen envelope.  The model in
which the helium shell is completely mixed with the hydrogen envelope
exhibits the most mixing. The general lack of internal mixing in PISNe
has several consequences for early chemical enrichment of the IGM, the
formation of second generation stars, and the observational signatures
of such explosions.

First, the elements that are expelled by low-mass Pop III SNe (which are 
governed by both mixing and fallback onto the compact remnant) are later 
imprinted on new stars in essentially the proportions in which they are 
created by the explosions, regardless of how and where the stars form.  This 
is because the metals are already highly mixed by the time the shock exits the 
star and are merely diluted upon further expansion into the halo and IGM, where 
new stars might form.  This implies that IMF averages of nucleosynthetic yields 
of primordial core-collapse SNe can be directly compared to chemical abundances 
in ancient metal-poor stars, without regard for any intervening hydrodynamical 
processes. Indeed, the fact that elemental yields from Salpeter IMF averages of 
15 - 40 M$_{\odot}$ progenitors in JET10 match those found in two observational 
surveys of extremely metal-poor (EMP) stars suggests that the bulk of early 
chemical enrichment may have been due to low-mass Pop III stars. This is at odds 
with the current state of the art in Pop III star formation simulations, which 
suggest that the first stars were predominantly 100 - 500 M$_{\odot}$.

Determining the nucleosynthetic imprint of PISNe on second-generation stars is 
much more problematic because their metals may have differentially contaminated 
new stars.  This is because metals in PISNe mixed with each other and with the 
surrounding IGM on much larger spatial scales, at radii where new stars may have
formed. Except in models u200 and u225 at the interface of the O and He shells, 
and in model u225 at the interface of the O and Si shells, the shells of elements 
in the ejecta of other PISNe expand homologously until they sweep up their own 
mass in the ambient H II region, at radii of 10 - 15 pc. At this point, a reverse 
shock detaches from the forward shock and a contact discontinuity forms between 
them \citep[the Chevalier phase--][]{wet08a}. Both the reverse shock and contact 
discontinuity are prone to dynamical instabilities that will mix elements from 
the interior of the remnant with the surrounding medium.  Later, on scales of 100 
- 200 pc, further mixing will occur upon collision of the remnant with the dense 
shell of the relic H II region. 

In this case, the elements that are taken up into new stars depends on how and 
where the stars form.  If metals migrate out into cosmological flows and then 
fall back into the the halo via accretion and mergers on timescales of 50 - 100 
Myr, they likely will be well-mixed, with all elements formed in the explosion 
appearing in the new star \citep{get07,wa08b,get10}.  However, new stars may form at 
much earlier times in the SN remnant at radii where mixing takes place. One way 
this could happen is if metals at the interface between the ejecta and swept-up 
shell suffuse into and cool the shell, causing it to fragment into clumps that 
are unstable to gravitational collapse \citep[e.g.][]{mbh03}. In this scenario, 
such clumps would be enriched only by the elements residing in the relatively 
narrow zone in which the stars form. This still does not explain why some hyper 
metal-poor stars have such high [C, N, and O/Fe] ratios. While C and O would be 
preferentially deposited on the clumps in which such stars form because they 
are predominant in the outer layers of the ejecta, PISNe do not produce enough 
N to account for measured abundances in these stars.  However, little iron from 
deep in the interior of the remnant would reach the clumps because it is not 
mixed, which is consistent with the abundances measured in EMP stars to date.
The failure to uncover the characteristic odd-even nucleosynthetic signature
of PISNe predicted by \citet{hw02} metal-poor star surveys thus far has led 
some to suppose that primordial stars may not have been very massive, but
it is possible that such signatures have been masked by observational selection
effects \citep{kjb09}. Unlike for low-mass stars, detailed numerical simulations 
that follow differential enrichment are required to determine the true chemical 
imprint of PISNe on second-generation stars.    

Second, our results imply that, unlike in SN 1987A or low-mass Pop III SNe, Ni 
and Fe emission lines will not appear immediately after radiation breakout from 
the shock because there is not enough mixing to transport these elements out to 
the photosphere of the fireball.  This may be key to distinguishing between Pop
III core-collapse explosions and PISNe. The light curves of these supernovae are 
characterized by a sharp intense initial transient that decays over several hours 
into a dimmer extended plateau that persists for 2 - 3 months in core-collapse 
events and 2 - 3 years in PISNe \citep{fwf10,wf10}.  The inital pulse is powered 
by the thermal energy of the shock and the plateau is energized by radioactive 
decay of \Ni\ (the long life of the plateau is due to the longer radiation 
diffusion timescales through the massive ejecta).  Preliminary radiation 
hydrodynamical calculations of PISN light curves and spectra indicate that their 
peak bolometric luminosities are similar to those of Type Ia and core-collapse 
SNe, making determination of the mass of the progenitor from the magnitude of the 
initial transient problematic.  If, however, Ni and Fe are detected just after 
radiation breakout, one can be confident that the explosion is due to a low-mass 
primordial star, and a Pop III IMF could begin to be built up by samples of such 
detections.  

Our models also demonstrate that one-dimensional radiation hydrodynamical models
are sufficient to capture most features of PISN light curves and spectra because
the mixing such calculations exclude, which would alter the order in which 
emission lines appear over time, is minor. The picture for core-collapse Pop III 
SNe is quite different because vigorous mixing prior to the eruption of the shock 
from the surface of the star mandates its inclusion in light curve models.  
Two-dimensional multigroup radiation hydrodynamical calculations of such spectra 
lie within the realm of current petascale platforms, but a less costly approach 
can incorporate mixing in one-dimensional models.  If most mixing in low-mass Pop 
III explosions occurs before radiation breaks free from the shock, two-dimensional 
models such as those in JET10 can be used to compute the distribution of elements 
in the ejecta just before breakout.  These explosions can then be azimuthally 
averaged onto the one-dimensional grid of the light curve calculation and evolved 
to compute spectra.  On average, along any given line of sight out of the SN, this
method will produce emission lines in the likely order they would be observed.

These simulations, together with our previous survey of mixing and fallback in 
low-mass Pop III SNe, are the first of a numerical campaign to model the chemical 
enrichment of the early cosmos from its smallest relevant spatial scales.  The 
eventual goal of this campaign is to understand the contribution of the first SNe 
to the formation of new stars and the assembly of primeval galaxies, which will 
soon be probed by the \textit{James Webb Space Telescope} (\textit{JWST}) and the 
\textit{Atacama Large Millimeter Array} (\textit{ALMA}).  The next stage of these 
numerical simulations will incorporate mixing on subparsec scales to determine if 
new stars directly form in the remnants of the first supernova explosions and 
follow their congregation into the first galaxies.

\acknowledgments

The authors thank Stan Woosley for helpful discussions and the use of
his \code{KEPLER} progenitor models. CCJ was supported in part by the
SciDAC Program under contract DE-FC02-06ER41438.  DJW acknowledges
support from the Bruce and Astrid McWilliams Center for Cosmology at
Carnegie Mellon University.  Work at LANL was done under the auspices
of the National Nuclear Security Administration of the U.S. Department
of Energy at Los Alamos National Laboratory under Contract
No. DE-AC52-06NA25396.  All simulations were performed on the open
cluster Coyote at Los Alamos National Laboratory.

\bibliographystyle{apj.bst}
\bibliography{refs.bib}

\end{document}